\documentclass[aas_macros]{ffclass}
\usepackage{graphicx}
\usepackage{natbib,stfloats}
\usepackage{mathrsfs}
\usepackage{url}
\usepackage{rotating}
\usepackage{hyperref}

\theoremstyle{TH}{

}

\theoremstyle{THrm}{

}

\theoremstyle{THhit}{

}

\makeatletter
\def\theequation{\arabic{equation}}

\makeatother

\begin{document}%
%%%%%%%%%%%%%%%%%

\setcounter{page}{1}

\LRH{G. K. Skinner et~al.}

\RRH{Science enabled by high precision inertial formation flying}

\VOL{x}

\ISSUE{x}

\PUBYEAR{xxxx}

\BottomCatch

%\CLline

\PUBYEAR{2013}

\subtitle{}

\title{Science enabled by high precision inertial formation flying}

\authorA{G. K. Skinner$^{1}$$^{*}$, B. R. Dennis$^{2}$,  \\ J. F. Krizmanic$^{2,3}$, E. P. Kontar$^{4}$}
\affA{$^{1}$ MPE, 85748 Garching, Germany \\
Phone: +49 89 30000 3828 \qquad E-mail: gerald.k.skinner@gmail.com\\
\vspace{2mm}
$^{2}$ NASA-GSFC, Greenbelt, MD 20770, USA\\
Phone: +1 301 286 7983/6817 \qquad E-mail: Brian.R.Dennis/John.F.Krizmanic@nasa.gov\\
\vspace{2mm}
$^{3}$ Universities Space Research Association (USRA), Columbia, MD 21044, USA\\
\vspace{2mm}
$^{4}$ Glasgow University, G12 8QQ, UK\\
Phone: +44 141 330 2499 \qquad E-mail: Eduard.Kontar@glasgow.ac.uk
\vspace{2mm}
 \newline
 $^{*}$Corresponding author }

%
%
%\authorB{B. Dennis}
%\affB{ NASA-GSFC, Greenbelt, MD 20770, USA\\
%Phone: +1 301 286 7983 \qquad E-mail: Brian.R.Dennis@nasa.gov}
%
%
%\authorA{Fusheng Wang\footnote{Work done while working at Siemens Corporate Research.} }
%
%\affA{Department of Biomedical Informatics, Emory University
%\newline
%36 Eagle Row, Ste 589, Atlanta, GA 30322, USA}
%
%
%
%\authorB{\footnotesize Cristobal Vergara-Niedermayr\footnote{Work done while working at Siemens Corporate Research.}}
%\affB{Oracle \newline
 %New Jersey, USA}
%
%

\begin{abstract}
The capability of maintaining two satellites in precise relative position, stable in a celestial coordinate system,
would enable major advances in a number of scientific disciplines and with a variety of types of instrumentation.
The common requirement is for formation flying of two spacecraft with the direction of their vector separation in inertial coordinates
precisely controlled and accurately determined as a function of time. We consider here the scientific goals that could be achieved
with such technology and review some of the proposals that have been made for specific  missions.
Types of instrumentation that will benefit from the development of this type of formation flying include 
1) imaging systems, in which an optical element on one spacecraft forms a distant image recorded by a detector array on the other spacecraft, including telescopes capable of very high angular resolution; 2) systems in which the front spacecraft of a pair carries an occulting disk, allowing very high dynamic range observations of the solar corona and exoplanets; 3) interferometers, another class of instrument that aims at very high angular resolution and which, though usually requiring more than two spacecraft, demands very much the same developments.
\end{abstract}

\KEYWORD{Astrophysics, Solar Physics, High angular resolution, Coronographs, Interferometry. }

%\REF{to this paper should be made as follows:
%Skinner, G. K., B. R. Dennis, J. F. Krizmanic, E. P. Kontar,
%`Science enabled by high precision inertial formation flying'
%(xxxx), {\it Int. J. of Space Science and Engineering},
%Vol. x, No. x, pp.xxx\textendash xxx.}

%\begin{bio}
%Manuel Pedro Rodr\'iguez Bol\'ivar received his PhD in Accounting at
%\end{bio}

\maketitle

 \section{Introduction}

Formation flying of two or more spacecraft could allow major advances in solar physics,  astronomy and astrophysics, but its potential is yet to be realised. Many missions have been suggested that depend on this technology, and some of them have been studied in detail. A payload on the Proba-3 technology demonstrator mission could be the first example of a class of  instruments that are much larger than an individual spacecraft.

This paper will concentrate on the science that would be enabled by the precision formation flying of at least two satellites
which are fixed in relative position in celestial coordinates, that is to say with respect to the `fixed' stars.
We will consider the general principles and review missions that have been studied and proposed that depend
on such technology. We will concentrate on missions involving long focal length telescopes
(\S2.1) and also coronagraphs and occulters (\S2.2), but also show that the same technology developments
would also make possible applications involving interferometers in various wavebands (\S2.3). We will then
review the technical challenges presented by the missions discussed and prospects for overcoming them (\S3).

\section{Studied/Proposed Precision Formation Flying Missions}
\subsection{Long telescopes}

Two considerations drive astronomers towards larger telescopes  - one is the need for a large collecting area
to collect enough photons from distant objects for imaging and/or spectroscopy.
The other is diffraction, which sets a limit on angular resolution, given by the Rayleigh criterion
\begin{equation}\label{eq1}
    \theta _{\text{diff}}\simeq 1.22 \frac{\lambda}{d} \;,
\end{equation}
for an  instrument with a circular aperture of diameter $d$ operating
at wavelength $\lambda$.
For bright sources (e.g. for solar and Earth-viewing instruments) the collecting area is often
not a driving requirement, but the second consideration remains.

In many circumstance large aperture ($d$) also means long focal length ($f$).
Whatever type of optics is used to converge the radiation towards a focus there will be some
limit to the angles $\phi$ through which it is practicable to redirect the radiation.
For the mirrors used in optical telescopes (or the `dishes' of radio telescopes) (Fig. 1a) the limit is only that set by off-axis aberrations
or by the cost of producing highly curved surfaces. Relatively low $f/d$ ratios can then be used.
Even for large $d$, focal lengths can then be modest and such designs are usually preferred
because short focal lengths minimise structural costs and are easier to implement.

In the X-ray part of the spectrum normal incidence reflecting optics cannot be used.
Reflecting optics then relies on the grazing incidence techniques, for which the angle between the arriving
radiation and the mirror surface must be less than a critical angle ${\alpha}_{max}$.
Depending on material, ${\alpha}_{max}$ is of the
order of ${1}^\circ$ for X-rays with $\lambda \sim 1$~nm (photon energy $E =1.24$~keV), and decreases
with increasing $E$. With the frequently-used Wolter-1 mirror configuration, two reflections are used
and the incoming radiation is redirected by $\phi=4\alpha$ (Fig. 2a).
As the angles involved are small $f =d/2\phi$${}_{max}$, so $f \geq d/8\alpha _{max}$.

Diffractive optics represent an alternative to reflective optics for X-ray imaging,
but again the angle through which the radiation can be redirected is limited. It is given by:
\begin{equation}\label{eq2}
    \phi =\frac{\lambda} {p}\simeq 4.2\left(\frac{1\,\mbox{keV}}{E}\right)\left(\frac{1\,\mbox{micron}}{p_x}\right) \simeq
    4.2\left(\frac{1\,\mbox{MeV}}{E}\right)\left(\frac{1\,\mbox{nm}}{p_{\gamma}}\right)\; \mbox{arcmin}\;,
\end{equation}
where $p_x$ is the period of the diffracting structure.
The first set of numerical values in Eq. (\ref{eq2}) is chosen to be typical of a situation in the X-ray band,
where the diffracting structure might be a micro-machined Zone Plate or Fresnel lens and $p_{x}$
is the period in a local region. The second one indicates approximate values that might
apply in the gamma-ray band, where Laue diffraction from crystals with an interatomic spacing $p_{\gamma}$
might be used. In all of these cases, because of  the need for a certain diameter to get an adequate
collecting area or angular resolution, the small $\phi$  drives one to using very long focal
lengths, $f$.

Two other considerations can also lead to the desirability of large $f$. First, in the X-ray and gamma-ray part of the spectrum
it is usually impossible to use secondary optics to magnify the primary image. Thus the size of a diffraction-limited image of a point source
is  $f \theta _{\text {diff}}$. Unless this is sufficiently large to be resolved by an available detector,
the full potential angular resolution cannot be realised.

Second, diffractive optics are chromatic - that is to say that the focal length depends on $\lambda$.
Consequently, only radiation with a specific wavelength will be perfectly focused on a detector
at a particular distance. Techniques exist to minimise the seriousness of this effect
\cite{2004SPIE.5488..601B, 2002A&A...383..352S,2003SPIE.4851..599G}, but they are effective only
over a limited band.  Large $f/d$ ratios minimise chromatic aberration.

Thus in the X-ray part of the spectrum, and {\it a fortiori} for gamma-rays, long focal lengths offer some major advantages.
Extensible structures have been used to allow instruments to be longer than can be accommodated within
a launcher fairing \citep{1994PASJ...46L..37T,2013ApJ...770..103H},
but of course there are limits to this approach. A number of instruments and missions have been proposed
which assume the use of formation flying of two spacecraft to allow even longer telescopes and
some have been the subject of extensive studies but none has yet reached the point of approval for flight.
Table 1 presents a list.  Some notes follow below.

\textbf{XEUS} is an X-ray observatory concept \citep{2009ExA....23..139A} studied by ESA in the context of the Cosmic Vision 2015-2025. It was
 designed to investigate the hot (million degree) Universe, from hot baryonic matter in the intergalactic medium
and the hot plasma in galaxy clusters to the immediate vicinities of black holes and of neutron stars where some of the most extreme
conditions in the Universe are found.

Use of grazing-angle reflecting optics based on Silicon micropore plates with a focal length of 35~m gave an effective area 40-70 times
greater than the current generation of instruments (ESA's XMM-Newton and NASA's Chandra). The table refers to the form in which {XEUS}
could have been placed in an L2 orbit with a single Ariane 5 launch. An earlier, larger, form required multiple launches and was limited
to operation in LEO, requiring a large amount of station keeping fuel, even for a limited life mission. The XEUS studies were merged
with NASA's Constellation-X studies, leading to the IXO proposal and in turn to Athena, neither of which involve formation flying.

\textbf{Simbol-X} was a joint CNES and ASI X-ray imaging mission planned to be launched in 2014 into a 4 day highly
elliptical orbit \citep{2009AIPC.1126....3F}. The design was based on the fact that  by utilizing either platinum coating
or multilayers, grazing incidence reflection can be used at relatively high energies. The grazing angles are very small, however.
By using a formation flying approach good performance was possible up to the energies of the 68~keV and 78~keV $^{44}$Sc lines from
the radioactive decay of $^{44}$Ti, an objective of prime importance for the astrophysics of supernovae explosions.
Unfortunately Simbol-X was abandoned in 2009 because of budgetary constraints.

\textbf{MAX}, \textbf{DUAL} and \textbf{GRI} are concepts that all have been proposed and studied to various levels and involve
optics using X-ray diffraction. Diffraction by an array of crystals in the Laue configuration allows X-rays and gamma-rays
to be concentrated onto a small, and hence low-background, detector. Laue lenses have been demonstrated
on the ground \citep{2004SPIE.5168..471H} and in a balloon flight \citep{2006NIMPA.567..333B} but not yet flown on a space mission.
Although their imaging properties are limited, the potential improvement in sensitivity over current technology is enormous, particularly for gamma-ray lines.
They offer the prospect of detecting gamma-ray lines from supernovae at distances up to 10-50~Mpc, enclosing a volume of space in which
several Type Ia supernovae occur every year. Understanding Type Ia supernovae is crucial to the distance scale
on which all cosmology theories depend.

\textbf{FLIP-3} was proposed to ESA in response to the 2012 call for `S' (=small) mission proposals. As it illustrates the issues associated
with using formation flying to obtain exceptionally high angular resolution, and as it is a mission with which some of the authors
of this paper were directly involved, we will consider it in some detail here.

One of the major goals of solar physics, and indeed of all astrophysics, is to understand how energy stored in cosmic magnetic fields
can be released explosively to produce such powerful phenomena as solar flares and coronal mass ejections and the even more powerful
outbursts from flare stars, accretion discs, etc. These fundamental processes of plasma energisation in solar flares drive the development
of new generation missions, e.g. SEE2020 \citep{SEE2020_white_paper} and SPARK \citep{2012ExA....33..237M}. Specifically, the hottest flare plasma,
with temperatures that can reach as high as 50 million degrees or even higher, is one of the results of this energy release,
and it can be most directly observed through its optically thin X-ray emission. It appears to occur through magnetic reconnection
in the ubiquitous magnetic loops observed before, during and after flares, but the structures are unresolved with present instrumentation.

The relative closeness of the Sun and the high photon fluxes measured in Earth orbit during a flare mean that relatively modest entrance apertures provide sufficient signal.
With cm-scale X-ray optics considerable improvement on the best angular resolution achieved to date ($\sim 1$ arc second) is theoretically
possible without encountering the diffraction limit.  By using diffractive optics, the technology exists to image on much finer scales than presently possible,
but relatively long focal lengths are required. An instrument design proposed by \cite{2012SoPh..279..573D} can improve
by at least an order of magnitude on the resolution that is currently the state of the art.

A formation flying architecture that meets all of the requirements of the instrument proposed by \cite{2012SoPh..279..573D} is in fact already
planned for the Proba-3 mission. Proba-3 will operate in a solar pointing mode with 100 m plus separation between the two spacecraft
and with precision adequate for X-ray flare imaging with 100 milli-arc-second resolution or better (Fig. \ref{fig:fig3}).
Consequently, when it appeared that the ASPIICS coronagraph (\cite{2009AdSpR..43.1007V} and \S2.3) would not be available as the science
 payload for the Proba-3 mission, a proposal \citep{2012_Flip3_proposal} was made to ESA in the context of the S-mission call to provide a solar
 flare X-ray imaging instrument to the Dennis et al. design. The FLIP3 instrument that was proposed consists simply of a small X-ray lens
 on the spacecraft closer to the Sun and a CCD detector array on the other spacecraft to record the image formed by the lens. The lens design can be
considered as a simplified Fresnel Lens or as a Phase Zone Plate and consists simply of a two level pattern of concentric rings etched into a silicon
substrate. Such lenses are routinely used with laboratory X-ray beams and the construction of large ones with long focal length,
that meet all the requirements for FLIP3, has been demonstrated at NASA-GSFC \citep{2012SoPh..279..573D}.
Existing CCDs and associated read-out electronics, similar for example to those used on Solar Dynamics Observatory \citep{2012SoPh..275....3P},
meet the detector requirements.

The requirements on the mission are to be able to maintain the relative positions of the two spacecraft such that the image
of a designated region of the Sun falls on the detector, for which 1 cm precision is adequate.
Retrospective knowledge of the alignment is rather more important ( $<0.5$~mm transverse displacement uncertainty)
and any drift of the alignment during a CCD integration needs to be minimised ($<0.05$~mm over 1-10 s). The distance between the spacecraft
needs to be held at the design focal length (nominally 100 m) but errors as large as 20~cm are acceptable without significant defocusing.
All of these requirements are met by the Proba-3 mission with ample margin. Fig. \ref{fig:fig4} shows a simulation of the sort of performance
that could be obtained with the FLIP-3 instrument.

The FLIP3 proposal did not survive the strong competition in the S-mission selection process and it now appears
that a form of the ASPIICS coronagraph will be flown after all. The FLIP-3 design concept remains, with all of the technologies ready,
awaiting a suitable flight opportunity.

The lens design proposed for FLIP-3 is one of the smallest, simplest, and shortest focal length members of a family of possible diffractive
optics X-ray and gamma-ray devices that have been discussed by various authors over the last 10
years \citep{2004SPIE.5488..601B,2007ApOpt..46.2586B,2003SPIE.4851..599G,2005SPIE.5900..369G,2002A&A...383..352S,2004ApOpt..43.4845S}.
All rely on the principle that focusing is achieved if it can be arranged that the phase of radiation that has passed through the device is the same
when it arrives at the focal point, irrespective of where over the entrance aperture it arrives.
To do this never requires the phase to be changed by more than $2\pi$. The refractive index of materials for X-rays
and gamma-rays is very close to unity. It is equal to $1-\delta$, where $\delta$  is a number that is small and decreases
with energy, typically as  $E^{-2}$. But the wavelength is also very small; consequently the thickness of material necessary
to achieve a $2\pi$  phase change is also small -- typically varying from a few microns in the X-ray band to a few mm at MeV energies. This
results in very lightweight optics.

\textbf{MASSIM} and \textbf{FRESNEL} are two mission concepts that illustrate the potential of long telescopes using formation flying
and diffractive optics to take astrophysics into entirely new realms.

The MASSIM design is based on six X-ray lenses, each an achromatic combination of a diffractive and a refractive component. The lenses are
 1~m in diameter and have overlapping passbands covering the energy range $4.5-11$~keV. MASSIM would be capable of milli-arc-second resolution
or better, allowing direct imaging of X-ray emission from the regions where astrophysical jets are accelerated and of  interacting winds
in binary systems.

MASSIM requires an inter-spacecraft distance of 1000~km and alignment precision and stability adequate to keep the image on detectors $\sim$100 mm in size.
Retrospective knowledge of drift changing the direction of the inter-spacecraft vector needs to be commensurate with the resolution.
1 milli-arc-second at 1000~km corresponds to 5~mm.

An even more extreme version of MASSIM has been studied with the objective of reaching the MICRO-arc-second angular resolution
that would be necessary to directly image the space-time around the event horizon of supermassive black holes such
as are found at the centres of active galaxies. In order that diffraction (Eq. 1) not be a limitation, it would be necessary to work in the gamma-ray
part of the spectrum. Fig. \ref{fig:fig5} shows a simulation of how a black hole might appear.
The FRESNEL concept would involve a lens $5-10$~m in diameter and a focal length of $10^5-10^6$~km.
Construction of the lens is remarkably  straightforward and
the detector array is within the capabilities of existing technology (the resolution requirement is $\sim1$~mm);
only the formation flying and associated navigation issues prevent the realisation of such a project.

\subsection{Coronagraphs and Occulting discs}

A second class of science missions demanding precision formation flying involve making observations
of a weak signal in the presence of much stronger flux of radiation. The ASPIICS coronagraph \cite{2009AdSpR..43.1007V}
designed for the Proba-3 mission indicates the principle. An occulting disc on the front spacecraft blocks direct radiation from the Sun
from falling on a telescope on the rear spacecraft, $\sim100$~m behind it (Fig. \ref{fig:fig3}).
This enables the observation of faint structures in the atmosphere just above the surface of the Sun.
Although the corona can be observed during rare eclipses by the moon and by instruments in which internal occulting discs are used,
an external occulter allows observations over long timescales and closer to the surface of the Sun and thus to the region in which the 
flare energy release from the coronal magnetic field is believed to take place.
%transition from temperatures of $\sim5000$~K to $\sim10^6$~K occurs. 
The ASPIICS instrument is designed to perform 5 arc-second
resolution imaging of the solar corona between $1.075$ and $3$ solar radii, with lower resolution observations down to $1.02$ solar radii.
As well as white light imaging, it has modes for imaging and spectrophotometry in the FeXIV and He~I~D$_3$ lines.

A more extreme example of the use of an occulter on a distant spacecraft is New Worlds Observer (NWO) \cite{2009SPIE.7436E...5C}
which would use a $50$~m diameter petal shaped occulter or `starshade' \cite{2006SPIE.6265E..55C}.
The starshade is designed to block light from a star allowing, like the TPF-I and Darwin interferometers discussed below,
much dimmer planetary bodies orbiting the star to be observed and characterised (Fig. \ref{fig:fig6}).
It requires a 4~m class telescope about 80000~km from the starshade but would be capable of a contrast limit of $10^{11}$
and observing as close as 50 milli-arc-seconds from the host star. One of the options that has been considered
is to use JWST as the telescope.

\subsection{Interferometers}

Extension of precision formation flying capability in celestial coordinates from two to several spacecraft allows an alternative approach
to very high angular resolution astronomical observations and to high contrast observations through interferometry.
It is interesting that the highest resolution currently available to astronomy is obtained by VLBI (Very Long Baseline Interferometry)
in the radio part of the spectrum where although $\lambda$ is long, the equivalent of the $d$ in Eq. 1
can be thousands of km and milli-arc second resolution is routinely obtained.
Although most radio observations can be made from the ground, there are reasons for going to space both to have very long baselines
and to operate at wavelengths beyond the extremes of the normal radio band.

Below about ten MHz, the ionosphere becomes opaque and astronomical observations cannot be made from the ground.
An instrument far from the Earth also has the advantage of minimising terrestrial interference.
A series of studies (ALFIS, ALFA, DARIS) have been made of interferometric systems based on a cluster of small satellites
and operating in the range $0.03-30$~MHz. Radio interferometers do not necessarily require rigid formations or precision formation flying
and can provide their own navigational information, deducing the locations of the receivers from their observations.
On the other hand, data handling and communication is a major concern. Thus the issues are rather different from those being
considered for direct imaging applications.

\textbf{TPF-I} and \textbf{Darwin}.  Like the NWO concept, infrared nulling interferometry is a way to observe
and characterize exo-planets and in particular potentially habitable ones. The infrared band is preferred over visible light because the ratio
of flux from a star to that from a planet orbiting it is less extreme. For an Earth-like planet and a Sun-like star it is,
however, still very large and removal of the flux from the star over a range of angular distances corresponding to what is considered
to be the `habitable zone' surrounding it is essential. Both NASA and ESA have studied infrared nulling interferometers
based on a cluster of spacecraft flying in a rigid formation. Both concepts, TPF-I 
\citep[NASA; ][]{2011SPIE.8151E..11M} and Darwin  \cite[ESA; ][]{2009ExA....23..435C} are based on several `collector' spacecraft
redirecting radiation to a common `combiner' spacecraft (Fig. \ref{fig:fig7}). CNES has also considered a simplified mission,
PEGASE, along the same lines \citep{2006SPIE.6265E..47L}.

Each TPF-I/Darwin `collector' spacecraft has a telescope with a mirror (up to $4.2$~m diameter in some versions studied)
whose pointing must be maintained to 10-50~milli-arc-sec. Even more challenging is that the $\sim100$~m optical paths between
spacecraft have to be controlled and stable at the level of a few nm, though this can be achieved by compensating somewhat
larger spacecraft displacements using variable optical delay lines.

\textbf{MAXIM}. It was noted in \S2.1 that micro-arc-second angular resolution is needed to directly image the region
around the event horizon surrounding super-massive black holes (SMBHs) and that in the gamma-ray band wavelengths
are so short that the diffraction limit allows this to be achieved with an optic $5-10$~m in diameter. Although gamma-rays
are known to be produced by SMBHs, the flux in X-ray photons is much greater and better understood.
However the diameter $d$ implied by Eq. 1 is then $100-1000$~m. A filled-aperture approach is impossible.
However, X-ray interferometry to achieve the same resolution is possible and has been considered \citep{2003ExA....16...91C}.

Like TPF-I and Darwin, the MAXIM proposal \cite{2005AdSpR..35..122C} requires multiple collector spacecraft
redirecting the incoming radiation towards one in which the beams are combined.
As discussed in \S2.1, X-rays can only be diverted through comparatively small angles,
and in addition the limited resolving power of detectors must be taken into account.
Consequently in this case the collecting spacecraft has to be far (e.g. $10-20$~km) behind the plane in which the collectors are distributed.
There are tight requirements on relative positions of the array of collector spacecraft and on their pointing,
but as for the FRESNEL concept the greatest challenge is the control, and particularly the knowledge,
of the pointing of the entire instrument, meaning in this case the direction, relative to the target,
of a line between a reference point in the collector array and the detector.
This must be determined with the micro-arc-second precision of the images sought.

As a variation on the MAXIM concept it has been pointed out that spacecraft with diffractive elements
could be used in place of the collectors \citep{2009ExA....27...61S}.
The tight requirements on the pointing and positioning of the collector spacecraft are relaxed (at the expense of a limited bandpass)
but the challenge of the overall pointing determination remains.

\section{Some Technical Aspects}

The missions reviewed above cover a wide range of scientific objectives and instrumental techniques but bringing them together indicates
how the formation flying requirements and the associated technical requirements often have much in common.
For astronomical observations there is always a requirement to hold a rigid formation fixed in a celestial coordinate system
(in the special case of solar observations there is a rotation once per year).
Although the precision with which this needs to be done is widely different in different cases,
there is also always a need to determine and control the orientation of the configuration to `point' at targets
of interest and to know where it is pointing.

The simplest case to consider is where there are just two spacecraft involved (Fig. \ref{fig:fig9}).
The instrument attitude is then the direction, relative to the direction of the target, of a `line of sight' joining
reference points on the two spacecraft. A pointing requirement translates into a constraint on the lateral
displacement ($\delta x$, $\delta y$, in Fig. \ref{fig:fig9}) of one spacecraft relative to a line passing
from a reference point on the other to the  target.

Although there may also be requirements on the attitude  ($\theta _x$, $\theta _y$) of the individual spacecraft,
on their roll angles, $\theta _z$, about the line of sight and on the distance $f$ separating the spacecraft,
these are usually secondary and less demanding. Considering, for example, the case of long telescopes,
such as FLIP-3, based on diffractive optics, errors in the attitude ($\delta\theta _x$, $\delta\theta _y$) of the individual spacecraft
only move the lens or the detector slightly out of a plane normal to the line of sight and have very little effect on the imaging.
In the case of the detector spacecraft, roll errors  $\delta\theta _z$ simply rotate the image.
For the lens spacecraft they have no effect at all.
Similarly, the defocusing caused by errors in the inter-spacecraft separation distance is small
because of the high focal ratio, $f/d$.

In the case of interferometers involving multiple spacecraft, constraints on the relative positions of all the components
and on the attitude of the spacecraft carrying collectors can be much tighter, but as for long telescopes, it is the celestial
orientation of the formation as whole that is crucial.

\subsection{Determination versus control}

It is important to distinguish between the need to control aspect and navigation parameters and  obtaining  measurements
to be used retrospectively for image reconstruction and data analysis. The alignment must be controlled well enough that,
for example, the important part of an image falls within the active detector area. Particularly at high energies the position,
energy and time of arrival of every photon is often recorded, so provided measurements are made that allow the precise
pointing at all times to be deduced retrospectively, blurring can be removed during data analysis.
In other cases a series of short integration `exposures' or `snapshots' are recorded, in which case it is important
that drifts during an exposure are negligible but longer term effects can be removed.

\subsection{Precision pointing determination}
Where the objective is to obtain very high angular resolution, determining the pointing can present a major challenge.
To take the most extreme examples discussed above,
to obtain (sub-) micro-arc-second astronomical imaging, (sub-) micro-arc-second pointing determination is needed.

Solar missions have the advantage of a bright ($-26.7^m$) reference star with very sharp limbs near the centre of the field of view.
Proba-3 will use a shadow position sensor, for example, to monitor the formation alignment with a lateral position uncertainty
expected to be only 15 microns \citep{2010SPIE.7731E..31L}.
Astronomical missions are likely to have to rely on stars, or possibly quasars.

To put the potential need for sub-micro-arc-second precision in context, the state-of-the-art in star-trackers is probably represented
by the JMAPS instrument (though the JMAPS mission was cancelled in 2012),
which was expected to achieve $\sim 5$~milli-arc-seconds (rms) single measurement uncertainty on stars down
to $12^m$ \citep{2010IAUS..261..350H}.  By the end of its 5-year life the GAIA astrometry mission to be launched in late 2013
should obtain parallax measurements with 10 micro-arc-second uncertainty \citep{2012AN....333..453P}.
For MAXIM the concept of a `super-startracker' using super sensitive gyroscopes has been discussed \citep{2004SPIE.5491..199C},
citing the performance demonstrated by Gravity Probe B.

\subsection{Station-keeping}
Maintaining the configuration will require a (pseudo-) continuous force on at least one spacecraft of each pair in order to overcome
disturbance forces -- notably gravity gradient effects, but also differences in radiation pressure and drag.
Gravity gradient forces on a configuration with a characteristic size $l$ in orbit at distance $r$ from a single central body
depend on $l/r^3$, so large $r$ are strongly favoured.
Of the systems discussed here, those with $l\leq 100$~m have generally been proposed for highly elliptical Earth orbits
(an exception being XEUS). Larger systems require operation in solar orbit far from the Earth or
at one of the Langrangian points.

As well as maintaining the formation, fuel is also generally needed for repointing to observe different targets.
If the fraction of the mission spent in repointing is not to be too large this can be a major consideration.
Fuel is always an issue but arguably a problem for which solutions exist. Some example studies are reported
by \cite{2005ExA....20..497K} and by \cite{2007SPIE.6687E..44L}.

Another issue that must be considered is that of the controllability of whatever thrusters are used.
Any deviation from precisely the force needed for station-keeping will lead to a rapidly building position error,
so fine control is needed while maintaining a relatively large continuous thrust.

%%%%%%%%%%%%%%%%%%%%%%%%%%%%%%%%%

\section{Conclusions}

A wealth of exciting science is possible if the formation flying capability needed for missions of the type discussed here can be developed.
Every mission mentioned here has been studied to at least a certain depth. In some cases there have been extensive studies and mission planning. None has yet been approved
for flight and most have been abandoned or placed on hold. Proposals founder on tests of affordability and credibility, particularly
because of doubts about the technical readiness of precise formation flying.

The missions described here under the heading of `long telescopes' all involve science instrumentation that can be built today
and could offer important scientific capabilities, including the possibility of some major advances in important fields.
The precision formation flying necessary to realize them involves not only holding the spacecraft in a rigid formation,
but holding that formation stable in a celestial frame of reference and determining its orientation with respect to that frame.
It appears that this should be possible, but the capability of doing so is yet unproven.
Ways need to be found to give credibility to our dreams.

%\section*{Acknowledgement}
%The project is funded in part by the National Institutes of Health, under Grant No. 5R01CA136535.

%%%%%%%%%%%%%%%%%%%%%%%%%%%%%%%%%%%%%%%%%%%%%%%%%%%%%%%%%%%%%%%%%%%%%%%%%%%%%%%%%%%%%

\bibliography{references}
\bibliographystyle{agsm5}

\begin{figure*}
\begin{center}
\includegraphics[trim = 25mm 115mm 25mm 15mm, clip, width=12cm]{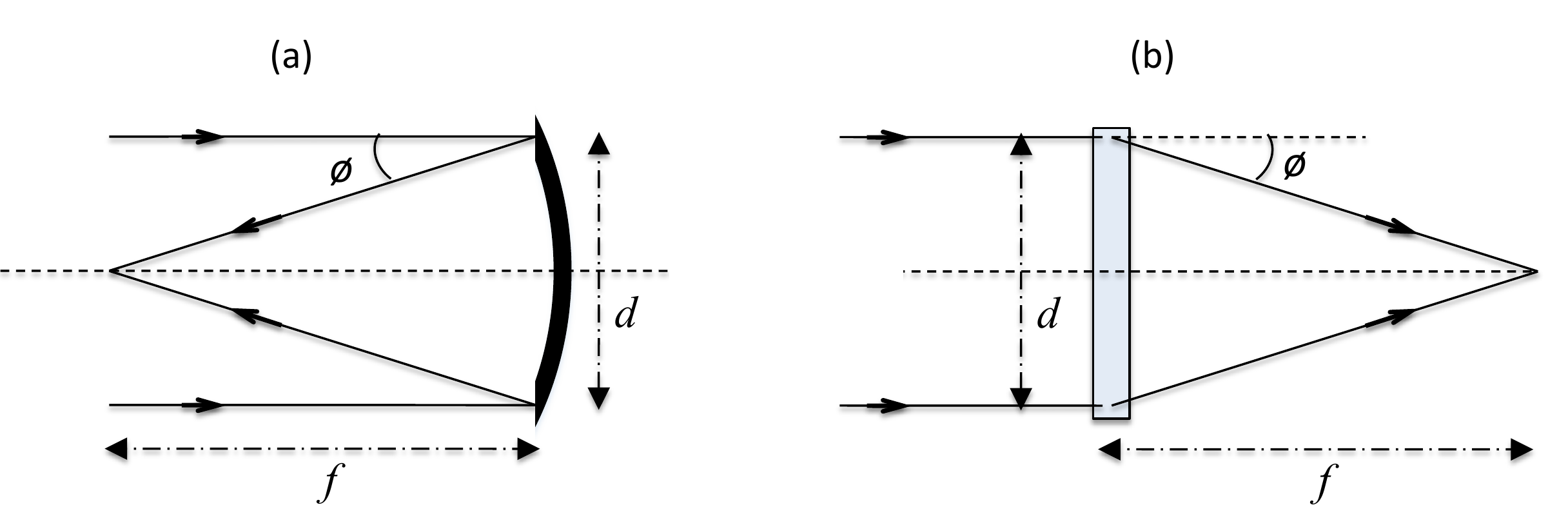}
\end{center}
\caption{  
(a) Illustrates the case of a reflecting telescope, in which the angle $\phi$  can be large so $f/d$  can be small (b) Shows an optical element used in ÔtransmissionÕ, which can be a lens, a grazing incidence mirror assembly or a diffractive element. }\label{fig:fig1}
%%\centerline{\epsffile{}}
\end{figure*}

\begin{figure*}
\begin{center}
\includegraphics[trim = 0mm 5mm 0mm 0mm, clip, width=12cm]{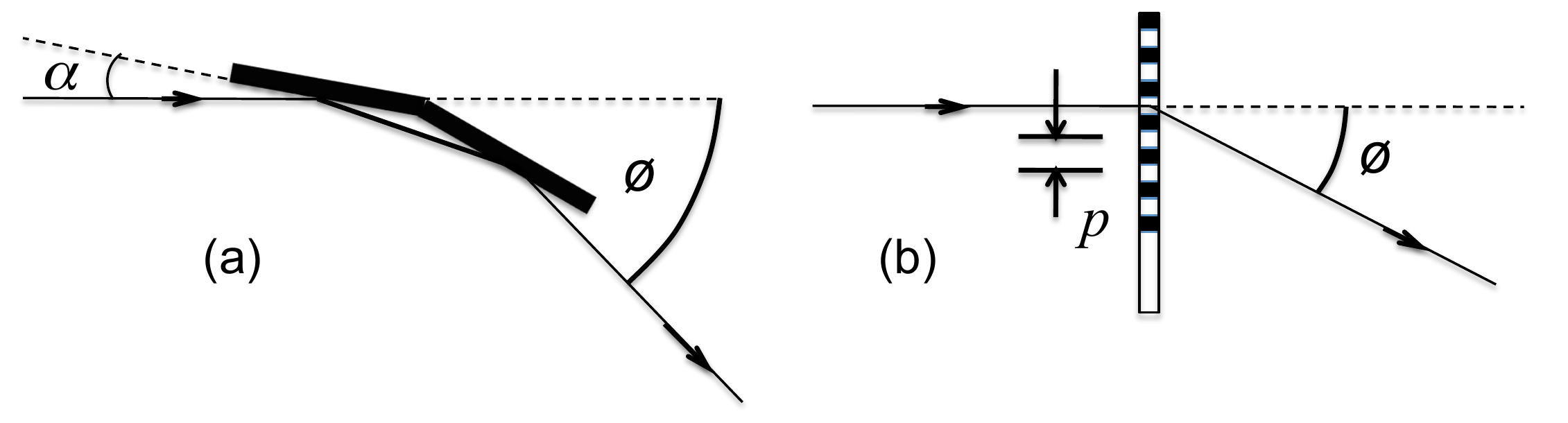}
\end{center}
\caption{For X-rays, the `transmissive' optical element in Fig. 1(b) can be (a) a grazing incidence mirror assembly
or (b) a diffractive element (Zone Plate, Phase Fresnel Lens or crystal).}\label{fig:fig2}
\end{figure*}

\begin{figure*}
\begin{center}
\includegraphics[trim = 50mm 100mm 60mm 70mm, angle=-90, clip, width=10cm]{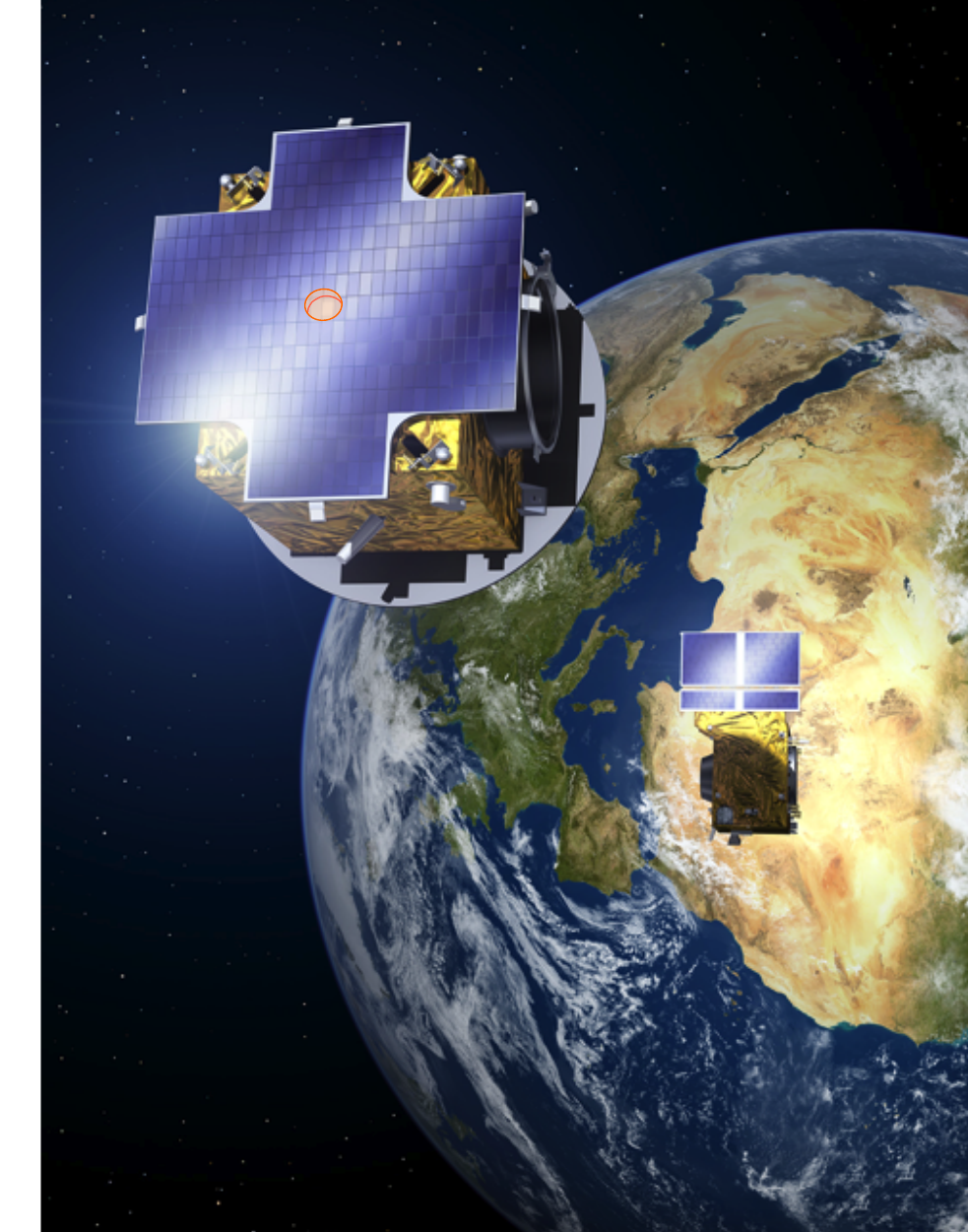}
\end{center}
\caption{The Proba-3 Formation Flying technology demonstration mission. Here a front spacecraft carries an occulting disc allowing instrumentation
on the rear spacecraft to make detailed observations of the solar corona \citep{2012_Flip3_proposal}.}\label{fig:fig3}
%%\centerline{\epsffile{}}
\end{figure*}

\begin{figure*}
\begin{center}
\includegraphics[angle=-90, trim = 0mm 10mm 0mm 0mm, clip, width=12cm]{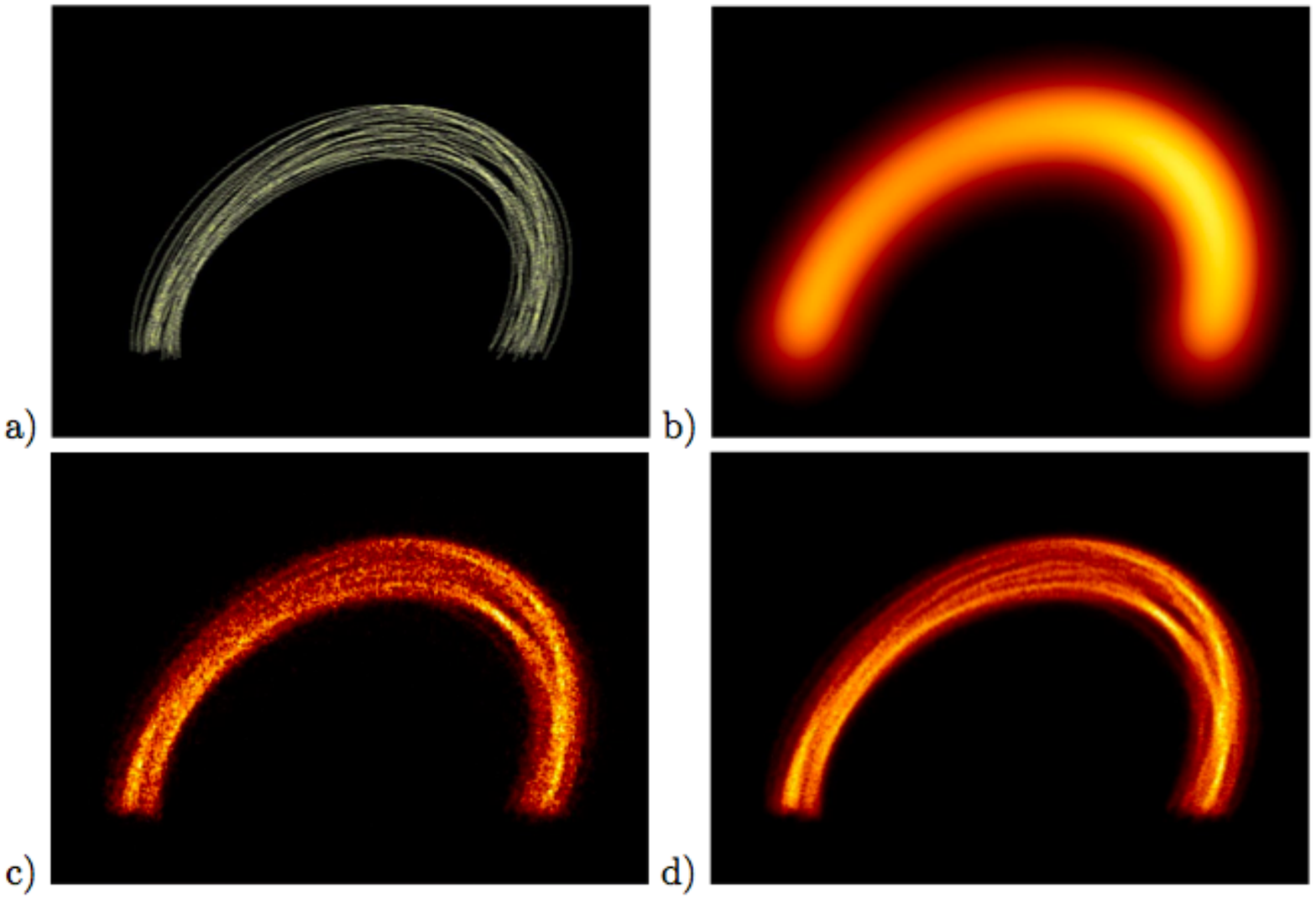}
\end{center}
\caption{Simulations of the FLIP-3 response to an assumed flare structure: (a) the assumed structure
(b) how this would be seen with the best currently available or planned angular resolution (1 arc sec)
(c) a 10 s snapshot with FLIP-3 (d) similar, but for a 10 times stronger flare or longer exposure.
The region shown is about 9 arc sec by 6 arc sec - only a small part of the 1.9 arc min square field of view.}
\label{fig:fig4}%%\centerline{\epsffile{}}
\end{figure*}

\begin{figure*}
\begin{center}
\includegraphics[trim = 0mm 0mm 0mm 0mm, clip, width=9cm]{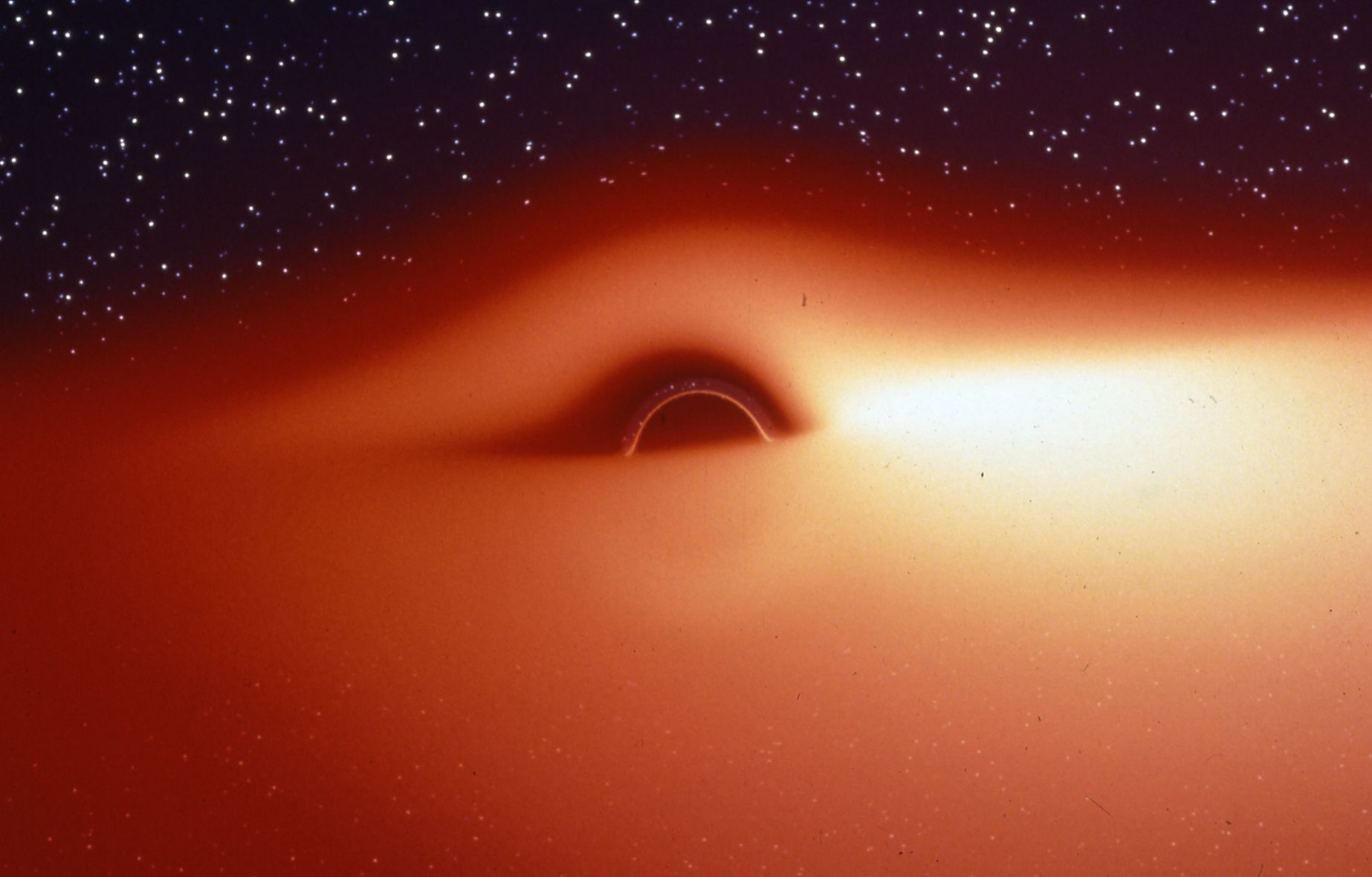}
\end{center}
\caption{Simulation of the appearance of a black hole surrounded by a thin accretion disk (J.-A. Marck).}
\label{fig:fig5}%%\centerline{\epsffile{}}
\end{figure*}

\begin{figure*}
\begin{center}
\includegraphics[trim = 0mm 0mm 0mm 0mm, clip, width=10cm]{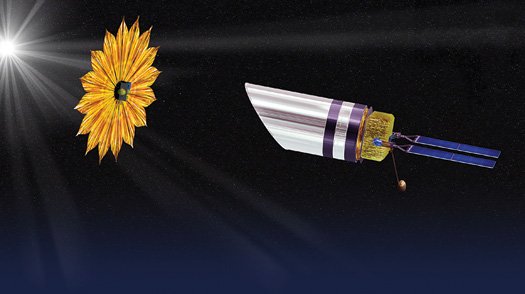}
\end{center}
\caption{The New Worlds Observer concept. An occulting disk with a petalled profile to minimize diffractive effects
blocks light from a star to enable observation of an exo-planet near to it  \citep{2009SPIE.7436E...5C}.}
\label{fig:fig6}%%\centerline{\epsffile{}}
\end{figure*}

\begin{figure*}
\begin{center}
\includegraphics[trim = 0mm 0mm 0mm 0mm, clip, width=10cm]{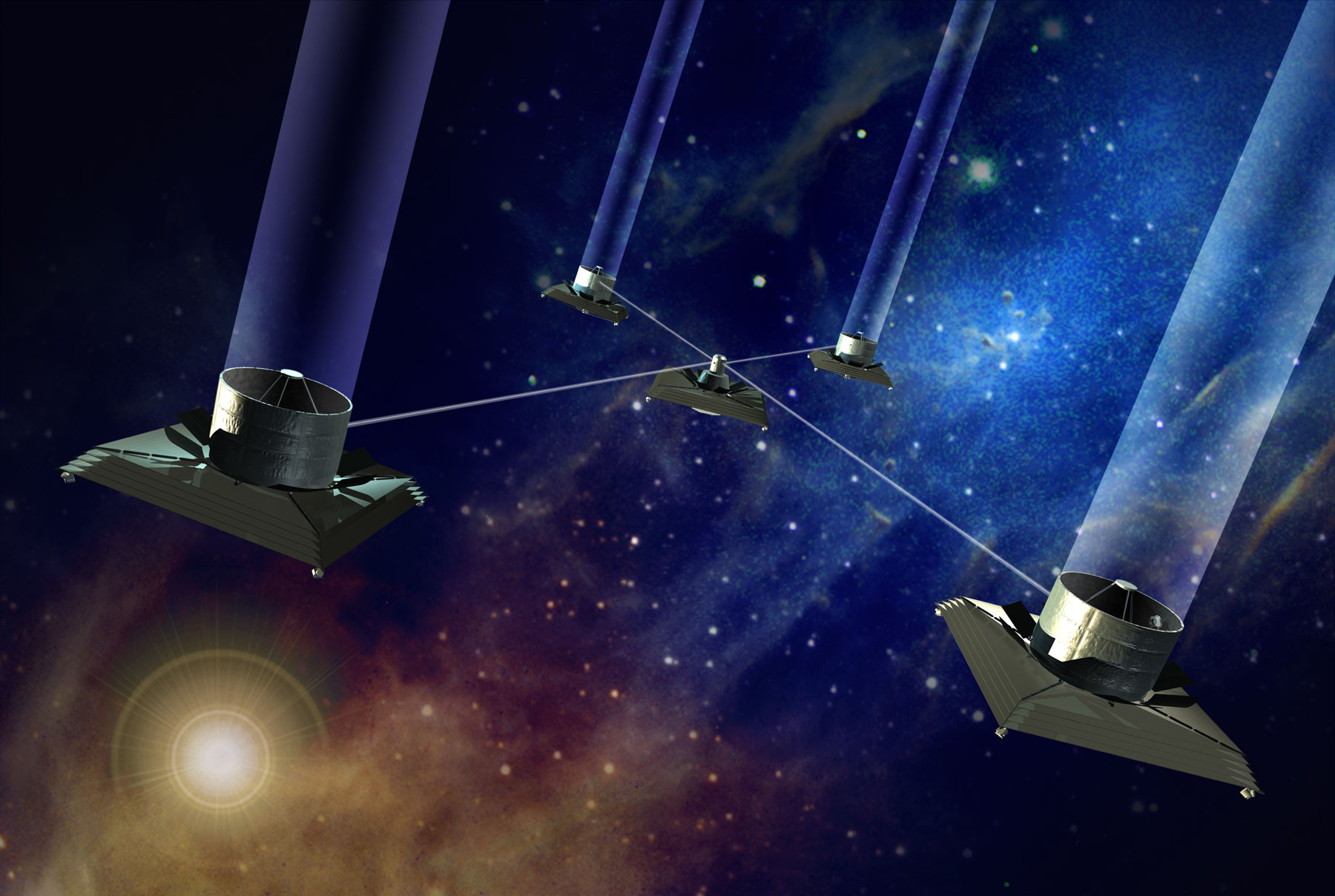}
\end{center}
\caption{The TPF-I concept. Free-flying `collector' spacecraft carrying IR telescopes direct radiation to a central `beam combiner' spacecraft \citep{TPFI.report}.}
\label{fig:fig7}%%\centerline{\epsffile{}}
\end{figure*}

\begin{figure*}
\begin{center}
\includegraphics[trim = 0mm 2mm 0mm 0mm, clip, width=10cm]{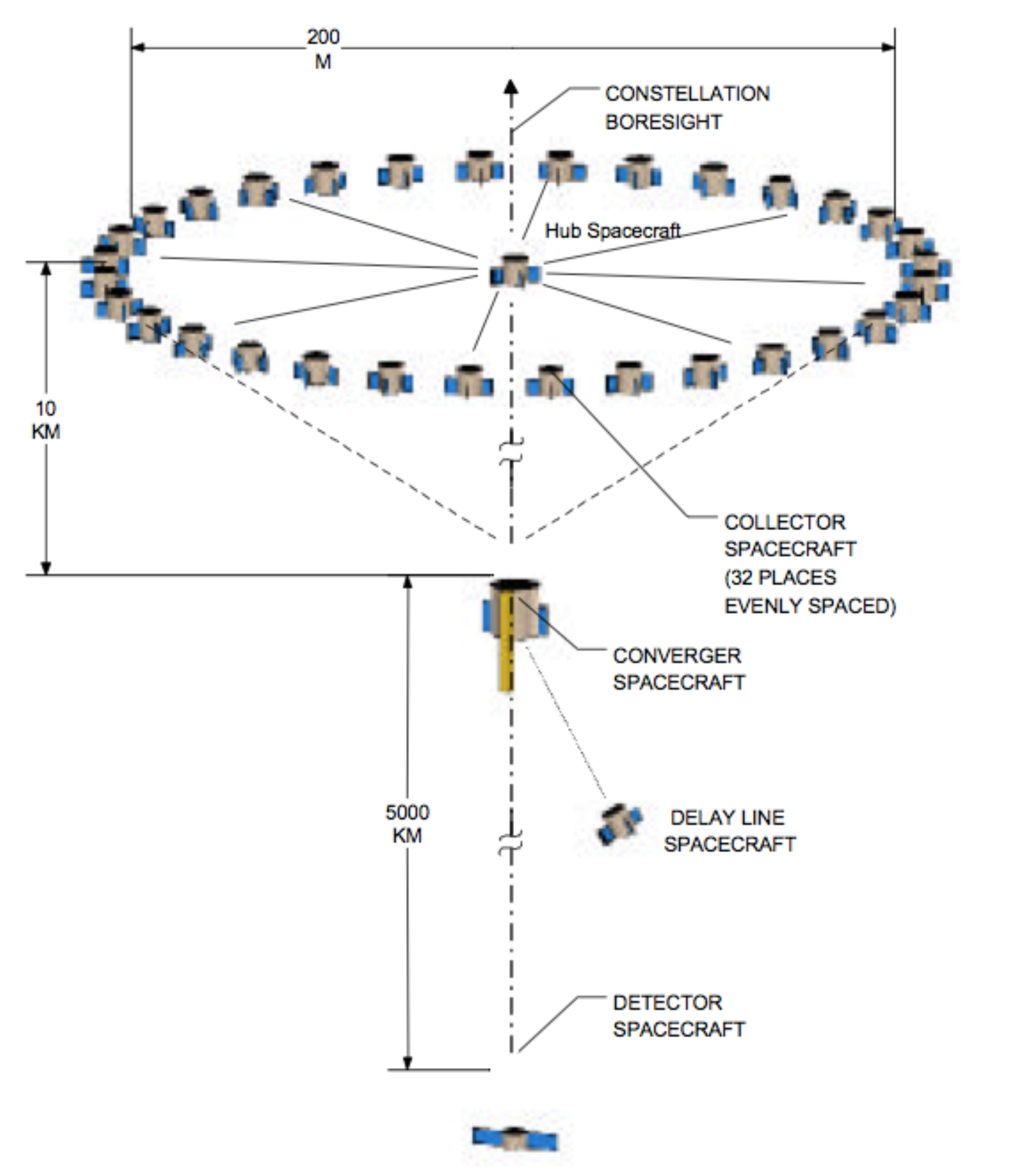}
\end{center}
\caption{X-ray interferometry with MAXIM. A ring of `collector' spacecraft, each of which  uses grazing incidence reflexion optics,  redirect X-rays towards a `converger' spacecraft \citep{2005AdSpR..35..122C}.}
\label{fig:fig8}%%\centerline{\epsffile{}}
\end{figure*}

\begin{figure*}
\begin{center}
\includegraphics[trim = 0mm 0mm 0mm 0mm, clip, width=10cm]{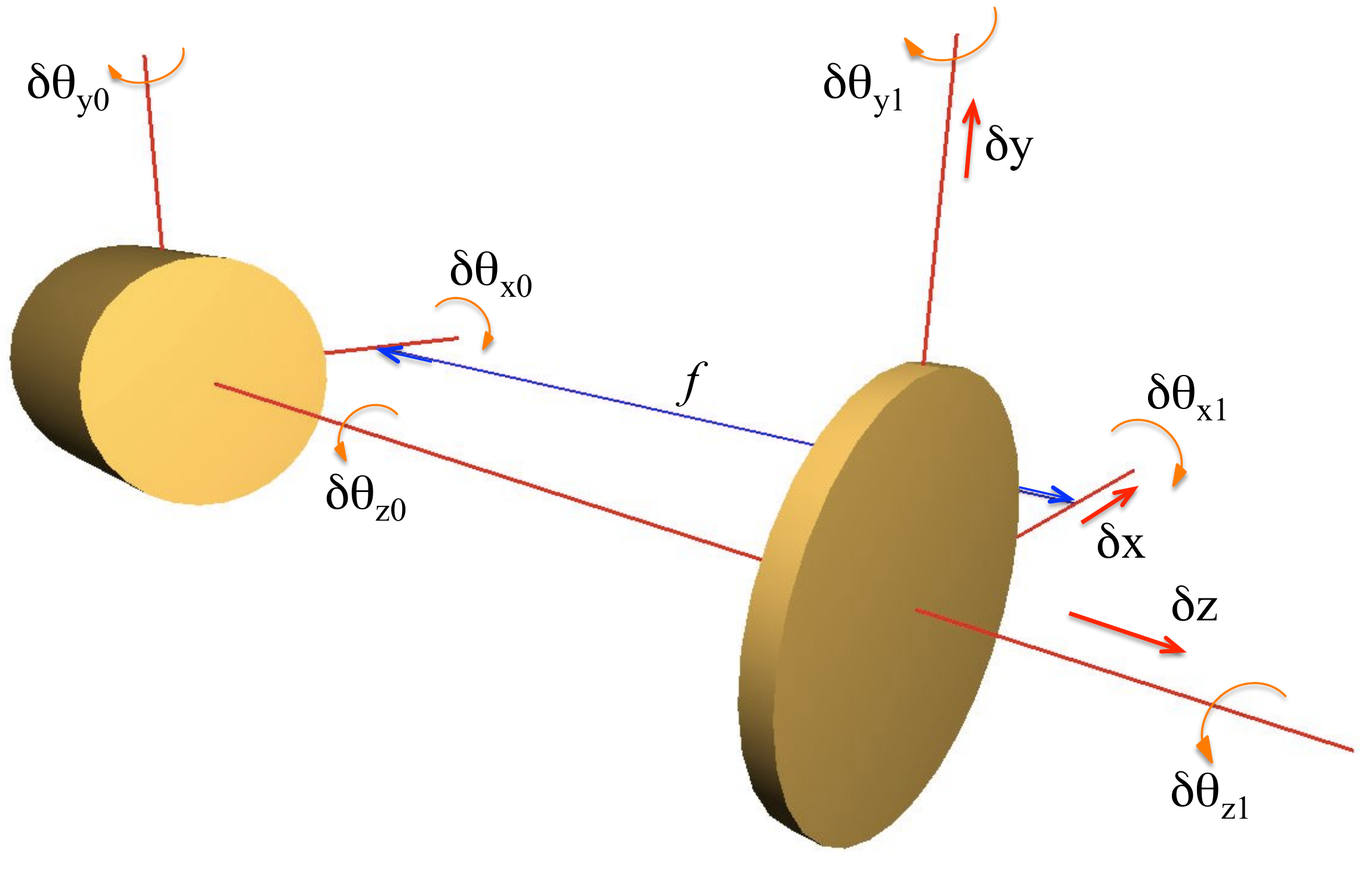}
\end{center}
\caption{For a long telescope or an occulting disc the most important requirement is usually on the offset, $\delta x$,
$\delta y$, of one satellite with respect to a line of sight from a reference point on the other spacecraft to the target.
The importance of the magnitudes and drifts of the other parameters indicated and on the uncertainties in their measurement
varies depending on the instrument.}
\label{fig:fig9}%%\centerline{\epsffile{}}
\end{figure*}

\begin{sidewaystable}
\caption{Studied/proposed missions using formation flying to form a long telescope. }
\begin{tabular}{|p{0.6in}|p{0.8in}|p{0.25in}|p{0.32in}|p{1.3in}|p{0.4in}|p{1.4in}|p{1.5in}|} \hline
\textbf{Mission} & \textbf{Energy (keV)} & \textbf{$d$(m)} & \textbf{$f$(m)} & \textbf{Optics} & \textbf{Angular Res.} (arcsec) & \textbf{Notes} & \textbf{References} \\ \hline
XEUS & 0.1--10 keV \break (40 keV extn.) & 4 &  35 & Grazing incidence \hfill \break  (Si micropore) &  5 & Led to IXO and Athena proposals & \cite{2009ExA....23..139A} \\ \hline
Simbol-X & 0.5--80 keV & 0.6 & 20-30 & Grazing incidence  \hfill \break (thin Ni shells) & 30 & CNES, ASI & \cite{2009AIPC.1126....3F} \\ \hline
MAX & 450--540 keV \hfill \break 800--900 keV & 2.2 & 86 & Laue Lens & $\sim60$ & CNES, SSL & \cite{2004ESASP.552..747V, 2005ExA....20..465B, 2006NIMPA.567..333B} \\ \hline
DUAL & 800-900 keV \hfill \break (0.1--10 MeV) & 2.2 & 68 & Laue Lens \hfill \break  ( \& Compton telescope) & $\sim60$ & JAXA, NASA, France \dots  & \cite{2006SPIE.6266E..60V,2012ExA....34..583V} \\ \hline
GRI & 50--2000 keV & 3.5 & 100 & Multi-layer grazing  \hfill \break  incidence + Laue Lens & 30 & ESA M class proposal & \cite{2007SPIE.6688E...5K,2009ExA....23..121K} \\ \hline
FLIP-3 & 6.7 keV & 0.01 & 100 & Phase Zone plate & 0.1 & Solar; ESA ``S'' mission proposal & \cite{2012SoPh..279..573D,2012_Flip3_proposal} \\ \hline
MASSIM & 4.5--11 keV & 1 (x6) & 10${}^{4}$ & X-ray Fresnel lenses  \hfill \break (achromatic) & 0.001 & NASA Advanced Mission Concept Study (proposed) & \cite{2008SPIE.7011E..22S}\\ \hline
FRESNEL & 500, 850 keV & 10 & 10${}^{9}$ & Gamma-ray Fresnel  \hfill \break lens & 10${}^{-6}$ & Studied at NASA-GSFC Mission Design Lab. 2001 & \cite{2003SPIE.4851.1366S} \\ \hline
\end{tabular}
\end{sidewaystable}

\end{document}